\documentstyle[preprint,aps,epsfig]{revtex}
\draft
\tighten

\begin{document}
\preprint{\vbox{\hbox{PNU-NTG-02/2000} \hbox{TPJU-5/2000}}}
\title{Spin structure of the octet baryons}
\author{Hyun-Chul Kim$^a$, Micha{\l} Prasza{\l}owicz$^b$, and Klaus Goeke$^c$}
\address{~\\
$^a$ Department of Physics, Pusan National University,\\
Pusan 609-735, Republic of Korea \\
$^b$ Institute of Physics, Jagellonian University, \\
ul. Reymonta 4, 30-059 Krak{\'o}w, Poland \\
$^c$ Institute for Theoretical Physics II, Ruhr-University Bochum, \\
D-44780 Bochum, Germany}
\date{\today}
\maketitle

\begin{abstract}
We analyze the semileptonic weak decays of the octet baryons in a 
"{\em model independent}" approach, based on the algebraic structure of
the Chiral Quark-Soliton Model. We argue that this analysis is in fact more
general than the model itself.  While the symmetry breaking for the 
semileptonic decays themselves is not strong, other quantities like 
$\Delta s$ and $\Delta \Sigma$ are much more affected.  We calculate 
$\Delta \Sigma$ and $\Delta q$ for all octet baryons.  Unfortunately, 
large experimental errors of $\Xi^-$ decays propagate in our analysis, 
in particular, in the case of $\Delta\Sigma $ and $\Delta s$.  Only if
the errors for these decays are reduced, the accurate theoretical predictions
for $\Delta\Sigma$ and $\Delta s$ will be possible.
\end{abstract}


\section{Introduction}

Since the European Muon Collaboration (EMC) measured the first moment of the
proton spin structure function $g_{1}^{p}$~\cite{EMC}, there has been a
great deal of discussion about the spin content of the proton. A series of
following 
experiments~\cite{SMC,E143,E154,HERMES1} 
confirmed the EMC measurement. In
contrast to the result from the naive nonrelativistic quark model, which is
reflected in the Ellis-Jaffe sum rule~\cite{EllisJaffe}, the strange quark
contribution to the nucleon spin deviates from zero. The global
fit performed by Ellis and Karliner \cite{EllKar} gives the following value $%
\Delta s=-0.11\pm 0.03$. For more recent analysis, see Refs.\cite{Cheng,Goto}.
These results, however, are obtained with an assumption
of the exact SU(3) symmetry for the baryon semileptonic decays.

Great efforts have been already spent on understanding the spin and flavor
content of the proton (see for review \cite{AEL} and
recent papers \cite{Cheng,Goto}). While it is now known that a large 
fraction of the nucleon spin is provided by gluons and their orbital angular 
momenta, it is still very important to understand the mechanism of 
how the quarks carry the nucleon spin. In particular, since the extraction 
of the flavor content of the nucleon spin relies on the empirical data of 
the baryon semileptonic decays, it is of great significance to examine the 
influence of the SU(3) symmetry breaking on the axial properties of the
baryons in a consistent way.

One piece of information comes from 
the first moment of the spin structure function $g_{1}^{{\rm {p}}}(x)$ of
the proton: 
\begin{equation}
I_{{\rm {p}}}=\int\limits_{0}^{1}dx\,g_{1}^{{\rm {p}}}(x)=\frac{1}{18}\left(
4\Delta u_{{\rm {p}}}+\Delta d_{{\rm {p}}}+\Delta s_{{\rm {p}}}\right)
\left( 1-\frac{\alpha _{{\rm {s}}}}{\pi }+\ldots \right) .  \label{Ip0}
\end{equation}
The analysis of Karliner and Lipkin~\cite{KarLip} implies 
$I_{{\rm {p}}}=0.124\pm 0.011$ which can be translated into: 
\begin{equation}
\Gamma _{{\rm {p}}}\equiv 4\Delta u_{{\rm {p}}}+\Delta d_{{\rm {p}}}+\Delta
s_{{\rm {p}}}=2.56\pm 0.23\,.  \label{Gampval}
\end{equation}
if $\alpha _{{\rm s}}(Q^{2}=3~({\rm GeV}/c)^{2})=0.4$ is assumed.  Let us 
for completeness quote also the result for the neutron:
\begin{equation}
\Gamma _{n}\equiv 
4\Delta d_{{\rm {p}}}+\Delta u_{{\rm {p}}}+\Delta
s_{{\rm {p}}} = -0.928\pm 0.186\,
\label{Gamnval}
\end{equation}
where the isospin symmetry (Bjorken sum rule) has been assumed.

Another piece of information comes from the semileptonic decays, which
in the case of the exact SU(3) symmetry can be parametrized by two
reduced matrix elements $F$ and $D$. Taking for 
$F=0.46$ and for $D=0.80$ together with Eq.(\ref{Gampval}), one gets for the
proton: $\Delta u_{{\rm {p}}}=0.79$, $\Delta d_{{\rm {p}}}=-0.47$ and $%
\Delta s_{{\rm {p}}}=-0.13$, which implies $\Delta \Sigma _{{\rm {p}}}=0.19$, 
quite a small number as compared with the naive expectation from the
quark model: $\Delta \Sigma _{{\rm {p}}}=1$.

It is important to realize that $\Delta \Sigma _{{\rm {p}}}$ is {\em not 
directly measured}; it is extracted from the data through some theoretical 
model. The standard way to calculate $\Delta \Sigma _{{\rm {p}}}$ is to assume
the SU(3) symmetry for the semileptonic decays. In this case it is enough to 
take {\em any} two decays and $\Gamma _{{\rm {p}}}$ of Eq.(\ref{Gampval}) as 
an input.
Normally, as in the example above, one uses neutron beta decay and 
$\Sigma^-$ decay as an input. However, if the SU(3) symmetry breaking was not 
important, any pair out of six known semileptonic decays should give roughly
the same number for $\Delta \Sigma _{{\rm {p}}}$. This is, however, {\em
not} the case. As we shall see in the next Section, 
$\Delta \Sigma _{{\rm {p}}}$
can be any number between 0.02 and 0.30. These numbers do not take into
account the experimental errors, therefore, as shown in Figure 1,
the uncertainty of $\Delta \Sigma _{{\rm {p}}}$ due to the SU(3)
symmetry breaking in the semileptonic decays is even larger. This is the
key observation which motivated this work.

It is almost impossible to analyze the symmetry breaking in weak decays
without resorting to some specific model~\cite{KarLip}. In this paper,
following Ref.~\cite{KimPraGo2}, we will implement the symmetry breaking for
the semileptonic decays using the Chiral Quark-Soliton Model
($\chi$QSM for short)~\cite{DPP,WakamatsuYoshiki} 
(see Ref.\cite{review} for review)
which satisfactorily describes the axial-vector properties of the hyperons 
\cite{BloPraGo}\nocite{BPG,Wakaspin}--\cite{KimPoPraGo}. 
Since the symmetry
breaking pattern of the $\chi$QSM is identical to the one derived in
large $N_{\rm c}$ QCD \cite{Man}, our analysis is in fact much
more general than the model itself.

However, since $g_{{\rm A}}^{(0)}($B$)$ does not correspond to the 
SU(3) octet axial-vector current, it is an independent quantity in QCD and 
it cannot be expressed in terms of $F$ and $D$ without some further 
assumptions. The $\chi$QSM 
(as most of the hedgehog models~\cite{SchechterWeigel}) has a remarkable 
virtue of connecting the singlet axial-vector constant with 
$g_{{\rm A}}^{(3)}$ and $g_{{\rm A}}^{(8)}$, and the semileptonic decay 
constants in a direct manner.  This connection introduces a model dependence 
into our analysis.  However, as we discussed in our previous paper on the 
proton  spin structure \cite{KimPraGo2} and as will be shown in Section V.A, 
there is no significant numerical difference between the results obtained 
with and without this model dependent ingredient. Whether this remains true 
for other baryons cannot be checked because of the lack of the data which 
could be additionally used if the model formula for $g_{{\rm A}}^{(0)}($B$)$ 
is abandoned.

In Section II.E we give an additional theoretical argument in favor of 
the model prediction for $g_{{\rm A}}^{(0)}($B$)$ .
 
In the previous paper \cite{KimPraGo2} we have shown how the symmetry breaking
influences the determination of $\Delta \Sigma _{{\rm {p}}}$ from the
existing data on the weak semileptonic baryon decays. Here our analysis is 
extended to all other members of the octet using the same "{\em %
model-independent}" method where the dynamical quantities,
which are in principle calculable within the model \cite{BloPraGo}, are
treated as free parameters. By adjusting them to the experimentally known
semileptonic decays we allow not only for maximal phenomenological input but
also for minimal model dependence. In 
Refs.\cite{KimPoPraGo}\nocite{strange,Min}--\cite{HongPark}
magnetic moments of the octet and decuplet have been studied in this way.
Model calculations for the vector-axial properties of baryons have been 
presented in Ref.\cite{KimPoPraGo}. There exist also direct model calculations 
of  the spin polarization function itself~\cite{WakaWata,Goekespin}.

Although the spin content of
the hyperons will be most probably not directly measured (with an exception of
$\Lambda$ where spin structure function can be related to the measured
fragmentation function \cite{BurkardtJaffe,Jaffe}), there is a substantial 
theoretical interest in the spin properties of the hyperons. 
We find that despite the fact 
that the symmetry breaking for the semileptonic decays themselves is not
strong, other quantities like $\Delta s$ and $\Delta \Sigma$ are much
more affected. We observe splitting of $\Delta\Sigma$ for different
baryons. Unfortunately our analysis suffers from large errors which
 are mainly due to the experimental errors of the $\Xi^-$ decays. 
It is therefore of utmost importance to measure these two decays 
with higher precision.

The paper is organized as follows: In Section II we recall the SU(3)
symmetry results and discuss various ways of determining $\Delta \Sigma$ 
and separately $\Delta q$'s. In Section III, following Ref.\cite{KimPraGo}, 
we recall 
the main properties of the $\chi$QSM with special emphasis on the mass
splittings, which we subsequently use in Section IV to parametrize the
SU(3) breaking of the semileptonic weak decays. In Section V numerical
analysis is carried out and the conclusions are given in Section VI.

\section{SU(3) symmetry at work}

Let us first briefly recall how the standard analysis is carried out. Three
diagonal axial-vector coupling constants define the integrated polarized quark
densities for a given baryon B: 
\begin{eqnarray}
g_{{\rm A}}^{(3)}({\rm {B}}) &=&\Delta u{\rm _{{B}}-\Delta d_{{B}},} 
\nonumber \\
\sqrt{3}g_{{\rm A}}^{(8)}({\rm {B}}) &=&{\rm \Delta u_{{B}}+\Delta d_{{B}%
}-2\Delta s_{{B}},}  \nonumber \\
g_{{\rm A}}^{(0)}({\rm {B}}) &=&{\rm \Delta u_{{B}}+\Delta d_{{B}}+\Delta s_{%
{B}}.}  \label{gA380}
\end{eqnarray}
Note that in our normalization $g_{{\rm A}}^{(0)}($B$)=$ $\Delta \Sigma _{%
{\rm {B}}}$.

Assuming the SU(3) symmetry, one can calculate $g_{{\rm {A}}}^{(3,8)}($B$)$ in
terms of the reduced matrix elements $F$ and $D$:\footnote{%
Note that $g_{{\rm {A}}}^{(3)}$ is proportional to $I_{3}$(third component
of the isospin which we assume to take the highest value )} 
\begin{eqnarray}
g_{{\rm A}}^{(3)}({\rm p})=F+D, & ~~~~& \sqrt{3}g_{A}^{(8)}({\rm p})=3F-D, 
\nonumber \\
g_{{\rm A}}^{(3)}(\Lambda )=0, & & \sqrt{3}g_{{\rm A}}^{(8)}(\Lambda )=-2D, 
\nonumber \\
g_{{\rm A}}^{(3)}(\Sigma ^{+})=2F, & & \sqrt{3}g_{{\rm A}}^{(8)}(\Sigma
^{+})=2D,  \nonumber \\
g_{{\rm A}}^{(3)}(\Xi ^{0})=F-D, & & \sqrt{3}g_{{\rm A}}^{(8)}(\Xi
^{0})=-3F-D.   \label{Eq:gA38}
\end{eqnarray}
At this stage $g_{{\rm {A}}}^{(0)}=\Delta \Sigma $ is an independent quantity 
and it is identical for all octet states. These equations together with 
(\ref{gA380}) allow one to express $%
\Delta q$'s in terms of $D$, $F$ and $\Delta \Sigma $:

\begin{eqnarray}
\Delta u_{{\rm {p}}}&=&1/3~\left( D+3F+\Delta \Sigma \right) ,  \nonumber \\
\Delta d_{{\rm {p}}}&=&1/3~\left( -2D+\Delta \Sigma \right) ,  \nonumber \\
\Delta s_{{\rm {p}}}&=&1/3~\left( D-3F+\Delta \Sigma \right) ,  \nonumber \\
\Delta u_{\Lambda }&=&1/3~\left( -D+\Delta \Sigma \right) ,  \nonumber \\
\Delta s_{\Lambda }&=&1/3~\left( 2D+\Delta \Sigma \right) ,  \nonumber \\
\Delta u_{\Sigma ^{0}}&=&1/3~\left( D+\Delta \Sigma \right) .   
\label{d}
\end{eqnarray}

The SU(3) symmetry imposes certain relations between $\Delta q$'s
of different flavor for different baryons:
\begin{eqnarray}
\Delta u_{{\rm {p}}} &=&\Delta u_{\Sigma ^{+}}=\Delta s_{\Xi ^{0}}, 
\nonumber \\
\Delta d_{{\rm {p}}} &=&\Delta s_{\Sigma ^{+}}=\Delta u_{\Xi ^{0}}, 
\nonumber \\
\Delta s_{{\rm {p}}} &=&\Delta d_{\Sigma ^{+}}=\Delta d_{\Xi ^{0}},
\label{Eq:SU3}
\end{eqnarray}
so that $\Delta q$'s given in Eq.(\ref{d}) are the  only 
independent ones in the SU(3) symmetry limit.
In addition we have the isospin relations 
\begin{eqnarray}
\Delta u_{{\rm {p}}} &=&\Delta d_{{\rm {n}}},\;\,\Delta d_{{\rm {p}}}=\Delta
u_{{\rm {n}}},\;\,\;\Delta s_{{\rm {p}}}=\Delta s_{{\rm {n}}},  \nonumber \\
\Delta u_{\Sigma ^{+}} &=&\Delta d_{\Sigma ^{-}},\,\Delta d_{\Sigma
^{+}}=\Delta u_{\Sigma ^{-}},\Delta u_{\Sigma ^{0}}=\Delta d_{\Sigma ^{0}} 
\nonumber \\
\Delta u_{\Lambda } &=&\Delta d_{\Lambda },\;\Delta s_{\Sigma ^{+}}=\,\Delta
s_{\Sigma ^{-}}=\Delta s_{\Sigma ^{0}},  \nonumber \\
\Delta u_{\Xi ^{0}} &=&\Delta d_{\Xi ^{-}},\,~\Delta d_{\Xi ^{0}}=\Delta
u_{\Xi ^{-}},\;\Delta s_{\Xi ^{0}}=\Delta s_{\Xi ^{-}}\qquad  \label{Eq:SU2}
\end{eqnarray}
which remain still valid after the inclusion of the SU(3) symmetry
breaking.

In order to find the numerical values of $\Delta q$'s one considers
different scenarios which we shortly discuss in the following.

\subsection{Naive quark model}

In the naive quark model there exist two relations between the constants $F$ 
and $D$ :
\begin{equation}
F/D=2/3,\;F+D=5/3 ~~\longrightarrow~~ F=2/3,\;D=1.  \label{SU3FD}
\end{equation}
Moreover, one assumes that the total spin is carried by the quarks, {\em %
i.e.}: 
\begin{equation}
\Delta \Sigma =1.
\end{equation}
With these parameters one gets $\Delta s_{{\rm {p}}}=0$. Values for all 
$\Delta q$'s and $\Gamma _{{\rm {p}}}$ are presented in Table I. The
prediction for $\Gamma _{{\rm {p}}}$ is, however, very bad, about two times
the experimental value.

\subsection{Extracting $F$ and $D$ from the semileptonic weak decays}

Certainly these naive quark model values (\ref{SU3FD}) are not realistic.
One can do better by extracting $F$ and $D$ from experiment. For example,
assuming the exact SU(3) symmetry, one has 
\begin{equation}
A_{1}\;=\;\left( g_{1}/f_{1}\right) ^{({\rm n}\rightarrow {\rm p}%
)}=F+D\,,\;\;\;\;A_{4}\;=\;\left( g_{1}/f_{1}\right) ^{(\Sigma
^{-}\rightarrow {\rm n})}=F-D\,.  \label{A1A4exp}
\end{equation}
For convenience, we denote the ratios of axial-vector to vector decay
constants by $A_{i}$ (see Table III). Taking for these decays the 
experimental values, one obtains 
\begin{equation}
F=0.46\;~~{\rm and}~~\;D=0.80,  \label{FDstand}
\end{equation}
as displayed in the column $(A_{1},\,A_{4})$ in Table I.

One could, however, use any two $A_{i}$'s out of six known weak semileptonic 
decays to extract $F$ and $D.$ The number of combinations is fourteen 
(actually fifteen, but two conditions are linearly dependent). 
Taking these fourteen combinations into account, one gets: 
\begin{equation}
F=0.40\div 0.55,\quad D=0.70\div 0.89\,.  \label{FDrange}
\end{equation}
These are the uncertainties of the {\em central values} due to the
theoretical error caused by using the exact SU(3) symmetry to describe the
weak semileptonic decays. These uncertainties are further increased by
the experimental errors of all individual decays.

Looking at Eq.(\ref{FDrange}), one might get an impression that a typical
error associated with the use of the SU(3) symmetry in analyzing the hyperon 
decays is of the order of 15 \% or so. While this is true for the hyperon 
decays themselves, the values of $\Delta q $ and $\Delta \Sigma $
for various baryons might be much more affected by the symmetry breaking.
Indeed, since
\begin{equation}
\Delta\Sigma =\frac{1}{2} \left(\Gamma_{\rm p} - 3 F - D \right)
\end{equation}
in the SU(3) symmetry limit we get
\begin{equation}
 \Delta \Sigma =0.02 \div 0.30
\end{equation} 
for $F$ and $D$ corresponding to Eq.(\ref{FDrange}) and $\Gamma _{{\rm {p%
}}}$ as given by Eq.(\ref{Gampval}).
This large uncertainty of the central value of $\Delta \Sigma $ is 
{\em entirely} due to the SU(3) symmetry breaking in the hyperon decays. In 
Fig.1 we plot  $\Delta \Sigma $ together with experimental errors for each 
pair of the semileptonic decays.

Anticipating the results of Section IV let us mention that there exist two
linear combinations of $A_{i}$'s which are free of the linear $m_{{\rm {s}}}$
corrections in the $\chi $QSM (and large $N_{\rm c}$ QCD \cite{Man}), namely: 
\begin{eqnarray}
F &=&\frac{1}{12}(4A_{1}-4A_{2}-3A_{3}+3A_{4}+3A_{5}+5A_{6}),  \nonumber \\
D &=&\frac{1}{12}(4A_{2}+3A_{3}-3A_{4}-3A_{5}+3A_{6})  \label{DFexp}
\end{eqnarray}
which give numerically
\begin{equation}
F=0.50\pm 0.07\;~~{\rm and}\; ~~ D=0.77\pm 0.04,  \label{FDmsfree}
\end{equation}
as displayed in Table I in the column ''average''. It is important to
note that by adopting this way of extracting  $F$ and $D$ in the
symmetry limit, no refitting of $F$ and $D$ is required when $m_{\rm s}$
corrections are added.

In what follows we shall
use these two sets -- Eqs.(\ref{FDstand},\ref{FDmsfree}) -- 
of values for $F$ and $D$ while discussing the predictions for $\Delta q$'s. 

In order to extract all $\Delta q$'s separately, one needs some additional 
information. Either another experimental input is needed, or a model which 
predicts $g_{\rm A}^{(0)}$ (${\rm B}$) in terms of $F$ and $D$.

\subsection{Conjecture of Ellis and Jaffe}

In 1974 Ellis and Jaffe \cite{EllisJaffe} made an assumption, based on the 
naive quark model that 
\begin{equation}
\Delta s_{{\rm {p}}}=0.
\end{equation}
From our SU(3) formula (\ref{d}), we see that this amounts to 
\begin{equation}
\Delta \Sigma =3F-D
\end{equation}
which indeed gives 1 for the naive quark model values (\ref{SU3FD}). For the
experimental values of $F$ and $D$ discussed in the previous section we get 
$\Delta \Sigma $ around 0.6 as displayed in Table I. Unfortunately, 
the value of $\Gamma_{\rm {p}}$ is much larger than the experimental value.

\subsection{Linking hyperon decays with the high energy data}

Instead of using the low energy data alone, one can also use the high energy
data on the first moment of the polarized structure function of the proton (%
\ref{Ip0}) with $\Gamma _{{\rm {p}}}=2.56$. The results of such fits for two
choices of $F$ and $D$ constants are presented in columns 5 and 6 of Table
I. A striking feature of these fits is that the resulting $\Delta \Sigma $
is very small. This fact is often referred to as a {\em spin crisis.}

\subsection{Chiral Quark Soliton Model}

As will be shown in the following, the $\chi $QSM predicts in the SU(3)
symmetry limit~\cite{BPG}: 
\begin{equation}
\Delta \Sigma =9F-5D  \label{ga0}
\end{equation}
for all octet baryons. This formula has a remarkable feature: It
interpolates between the naive quark model and the Skyrme model. 
Indeed, for
(\ref{SU3FD}) $\Delta \Sigma =1$, whereas in the case of the simplest Skyrme
model for which $F/D=5/9$, $\Delta \Sigma =0$, as observed for the first
time in Ref.\cite{BroEllKar}.

Here $\Delta \Sigma $ is very sensitive to small variations of $F$ and $D$,
since it is a difference of the two, with relatively large coefficients.
Indeed, for the 14 fits mentioned before Eq.(\ref{FDrange}) the central 
value for $\Delta \Sigma $ varies between $-0.25$ to approximately 1. 
Thus, despite the fact that the hyperon semileptonic decays are relatively 
well described by the model in
the SU(3) symmetry limit, the singlet axial-vector constant is basically
undetermined. This is a clear signal of the importance of the symmetry
breaking for this quantity.

In fact, conclusions similar to ours have been obtained in chiral
perturbation theory in Ref.\cite{SavWal}.

\section{Mass splittings in the $\protect\chi $QSM}

In this Section we shall briefly recall  how the model
parameters are fixed. Because of the SU(3) symmetry breaking due to the
strange quark mass $m_{{\rm s}}$ the collective baryon Hamiltonian is no
longer SU(3)-symmetric. Indeed \cite{Blotzetal}: 
\begin{equation}
H=H_{0}+H^{\prime }  \label{HSU3}
\end{equation}
where 
\begin{equation}
H_{0}=M_{{\rm {sol}}}+\frac{1}{2I_{1}}S(S+1)+\frac{1}{2I_{2}}\left( C_{2}(%
{\rm {SU(3)}}-S(S+1)-\frac{N_{{c}}^{2}}{12}\right)   \label{H0}
\end{equation}
and 
\begin{equation}
\hat{H^{\prime }}=m_{{\rm s}}\left( \alpha D_{88}^{(8)}+\beta \hat{Y}+\frac{%
\gamma }{\sqrt{3}}\sum\limits_{A=1}^{3}D_{8A}^{(8)}\hat{S}_{A}\right) .
\label{Hprim}
\end{equation}
Here $\hat{S}_{A}$ denotes baryon spin, $C_{2}($SU(3)$)$ 
the Casimir operator and $%
D_{B\,S}^{({\cal R})}$ are the SU(3) Wigner matrices in representation 
${\cal R}$.
Constants $\alpha $, $\beta $ and $\gamma $ are given by Ref.\cite{Blotzetal}: 
\begin{equation}
\alpha =-\sigma +\frac{K_{2}}{I_{2}},~~~~\beta =-\frac{K_{2}}{I_{2}}%
,~~~~\gamma =2\left( \frac{K_{1}}{I_{1}}-\frac{K_{2}}{I_{2}}\right) .
\label{albega}
\end{equation}
Here $K_{i}$ and $I_{i}$ are the ``moments of inertia'' and $\sigma $ is
related to the nucleon sigma term: $3 \sigma = \Sigma /%
\overline{m}$, $\overline{m}$ being the average mass of the up and down quarks.

The collective splitting Hamiltonian (\ref{Hprim}) mixes the states in
various SU(3) representations. The octet states are mixed with the higher
representations such as antidecuplet $\overline{{\bf 10}}$ and
eikosiheptaplet ${\bf 27}$. In the linear order in $m_{{\rm s}}$ the wave
function of a state $B=(Y,I,I_{3})$ of spin $S_{3}$ is given as: 
\begin{equation}
\psi _{B,S_{3}}=(-)^{\frac{1}{2}-S_{3}}\left( \sqrt{8}\,D_{B%
\,S}^{(8)\;\ast}+c_{B}^{(\overline{10})}\sqrt{10}\,
D_{B\,S}^{(\overline{10})\;\ast}
+c_{B}^{(27)}\sqrt{27}\,D_{B\,S}^{(27)\;\ast}\right) ,  \label{wavef}
\end{equation}
where $S=(-1,\frac{1}{2},S_{3})$. Mixing parameters $c_{B}^{({\cal R})}$ can
be found for example in Ref.~\cite{BloPraGo}. They are given as products of
$m_{{\rm s}}$ (which we assume to be 180~MeV) times a known numerical constant 
$N_{B}^{({\cal R})}$ depending on the baryonic state $B$
and a dynamical parameter $c_{{\cal R}}$. Since $c_{{\cal R}}$ depends on
the model parameter $I_{2}$, which is
responsible for the splitting between the octet and higher exotic multiplets~
\cite{antidec} and is not constrained from the data we will take them
as free parameters in our fits.

\section{Semileptonic weak decays in the chiral quark-soliton model}

The transition matrix elements of the hadronic axial-vector current 
$\langle B_2 |A_\mu^X | B_1\rangle$ can be expressed in terms of three 
independent form factors: 
\begin{equation}
\langle B_2| A_\mu^X |B_1\rangle\;=\;\bar{u}_{B_2} (p_2) \left[ \left\{g_1
(q^2) \gamma_\mu - \frac{i g_2 (q^2)}{M_1} \sigma_{\mu\nu} q^\nu + \frac{g_3
(q^2)}{M_1} q_\mu\right\}\gamma_5 \right] u_{B_1} (p_1),
\end{equation}
where the axial-vector current is defined as 
\begin{equation}
A_{\mu}^X\;=\; \bar{\psi}(x) \gamma_\mu \gamma_5 \lambda_X \psi (x)
\label{Eq:current}
\end{equation}
with $X= \frac12 (1 \pm i 2)$ for strangeness conserving $\Delta S = 0$
currents and $X=\frac12 (4 \pm i 5) $ for $|\Delta S| = 1 $.

The $q^{2}=-Q^{2}$ stands for the square of the momentum transfer $%
q=p_{2}-p_{1}$. The form factors $g_{i}$ are real quantities depending only
on the square of the momentum transfer in the case of $CP$-invariant
processes. We can safely neglect $g_{3}$ for the reason that on account of $%
q_{\mu }$ its contribution to the decay rate is proportional to the ratio $%
\frac{m_{l}^{2}}{M_{1}^{2}}\ll 1$, where $m_{l}$ represents the mass of the
lepton ($e$ or $\mu $) in the final state and $M_{1}$ that of the baryon in
the initial state. Similarly we shall neglect $g_2$. In principle this form
factor is proportional to $m_{\rm s}$ and therefore should be included in
the consistent analysis of the weak decays data. Unfortunately, such an
analysis is still missing and all experimental results on $g_1$ assume
$g_2 \equiv 0 $.

Another possible small $m_{\rm s}$ corrections come from the evolution
of $g_1$ with $Q^2$, due to the non-conservation of the
axial-vector currents caused by the SU(3) symmetry breaking. These
corrections are also neglected in our approach.

It is already well known how to treat hadronic matrix
elements such as $\langle B_{2}|A_{\mu }^{X}|B_{1}\rangle $ in the $\chi 
$QSM (see for example \cite{review} and references therein). Taking into
account the $1/N_{c}$ rotational and $m_{{\rm s}}$ corrections, we can write
the resulting axial-vector constants $g_{1}^{B_{1}\rightarrow B_{2}}(0)$ in the
following form: 
\begin{eqnarray}
g_{1}^{(B_{1}\rightarrow B_{2})} &=&a_{1}\langle
B_{2}|D_{X3}^{(8)}|B_{1}\rangle \;+\;a_{2}d_{pq3}\langle B_{2}|D_{Xp}^{(8)}\,%
\hat{S}_{q}|B_{1}\rangle \;+\;\frac{a_{3}}{\sqrt{3}}\langle
B_{2}|D_{X8}^{(8)}\,\hat{S}_{3}|B_{1}\rangle  \nonumber \\
&+&m_{s}\left[ \frac{a_{4}}{\sqrt{3}}d_{pq3}\langle
B_{2}|D_{Xp}^{(8)}\,D_{8q}^{(8)}|B_{1}\rangle +a_{5}\langle B_{2}|\left(
D_{X3}^{(8)}\,D_{88}^{(8)}+D_{X8}^{(8)}\,D_{83}^{(8)}\right) |B_{1}\rangle
\right.  \nonumber \\
&+&\left. a_{6}\langle B_{2}|\left(
D_{X3}^{(8)}\,D_{88}^{(8)}-D_{X8}^{(8)}\,D_{83}^{(8)}\right) |B_{1}\rangle 
\right] ,  \label{Eq:g1}
\end{eqnarray}
where $a_{i}$ denote parameters depending on the specific dynamics of the
chiral soliton model. Their explicit form in the $\chi $QSM can be found in
Ref.~\cite{BloPraGo}. 

Analogously to Eq.(\ref{Eq:g1}) one defines the diagonal axial-vector 
couplings. In that case $X$ can take two values: $X=3$ and $X=8$. For $X=0$ 
(singlet axial-vector current) we have the following expression 
\cite{BloPraGo,BPG}: 
\begin{equation}  \label{Eq:singlet}
g_B^{(0)}\, \hat{S}_3 = a_3\, \hat{S}_3 + \sqrt{3} m_{\rm s}(a_5-a_6) \; 
\langle B| D^{(8)}_{83} | B \rangle .
\end{equation}

A remark concerning constants $a_i$ is here in order. Coefficient $a_1$
contains the terms which are leading and subleading in the $1/N_{{\rm c}}$
expansion. The presence of the subleading terms enhances the numerical value
of $a_1$ calculated in the $\chi$QSM for the self-consistent profile and makes 
the model predictions remarkably close to the experimental data 
\cite{WaWa,allstars}. 
This feature, although very important for the model phenomenology, does not
concern us here, since our procedure is based on fitting all coefficients $%
a_i $ from the data. Constants $a_2$ and $a_3$ are both subleading in $1/N_{%
{\rm c}}$ and come from the anomalous part of the effective Euclidean
action. In the Skyrme model they are related to the Wess-Zumino term.
However in the simplest version of the Skyrme model (which is based on the
pseudo-scalar mesons only) $a_3 = 0$ identically \cite{BroEllKar}. In the
case of the $\chi$QSM $a_3 \ne 0$ and it provides a link between the SU(3) 
octet of axial-vector currents and the singlet current of 
Eq.(\ref{Eq:singlet}).  It was
shown in Ref.\cite{limit} that in the limit of the artificially large soliton,
which corresponds to the ``Skyrme limit'' of the present model, $a_3/a_1
\rightarrow 0$ in agreement with \cite{BroEllKar}. On the contrary, for the
small solitons $g_{\rm p}^{(0)} \rightarrow 1$ reproducing the result of 
the non-relativistic quark model.

Instead of calculating 7 dynamical parameters $a_{i}$ and $I_{2}$ 
(or $c_{\overline{10}}$ and $c_{27}$) within the $\chi $QSM
(which was done in Ref.\cite{KimPoPraGo}, we shall fit them from the weak 
semileptonic decay data. It is convenient to introduce the following set of new
parameters: 
\[
r=\frac{1}{30}\left( a_{1}-\frac{1}{2}a_{2}\right) ,\;\;\;\;\;\;s=\frac{1}{60%
}a_{3},\;\;\;x=\frac{1}{540}m_{{\rm s}}\,a_{4},\;\;\;y=\frac{1}{90}m_{{\rm s}%
}\,a_{5},\;\;\;z=\frac{1}{30}m_{{\rm s}}\,a_{6}, 
\]
\begin{equation}
p=\frac{1}{6}m_{{\rm s}}\,c_{\overline{10}}\left( a_{1}+a_{2}+\frac{1}{2}%
a_{3}\right) ,\;\;\;q=-\frac{1}{90}m_{{\rm s}}\,c_{27}\left( a_{1}+2a_{2}-%
\frac{3}{2}a_{3}\right) .\label{Eq:newp}
\end{equation}

Employing this new set of parameters, we immediately express all possible
semileptonic decay constants between the octet baryons: 
\begin{eqnarray}
\left( {g_{1}}/{f_{1}}\right) ^{({\rm n}\rightarrow {\rm p})}
&=&-14r+2s-44x-20y-4z-4p+8q,  \nonumber \\
\left( {g_{1}}/{f_{1}}\right) ^{(\Sigma ^{+}\rightarrow \Lambda )}
&=&-9r-3s-42x-6y-3p+15q,  \nonumber \\
\left( {g_{1}}/{f_{1}}\right) ^{(\Lambda \rightarrow {\rm p})}
&=&-8r+4s+24x-2z+2p-6q,  \nonumber \\
\left( {g_{1}}/{f_{1}}\right) ^{(\Sigma ^{-}\rightarrow {\rm n})}
&=&4r+8s-4x-4y+2z+4q,  \nonumber \\
\left( {g_{1}}/{f_{1}}\right) ^{(\Xi ^{-}\rightarrow \Lambda )}
&=&-2r+6s-6x+6y-2z+6q,  \nonumber \\
\left( {g_{1}}/{f_{1}}\right) ^{(\Xi ^{-}\rightarrow \Sigma ^{0})}
&=&-14r+2s+22x+10y+2z+2p-4q,  \nonumber \\
\left( {g_{1}}/{f_{1}}\right) ^{(\Sigma ^{-}\rightarrow \Lambda )}
&=&-9r-3s-42x-6y-3p+15q,  \nonumber \\
\left( {g_{1}}/{f_{1}}\right) ^{(\Sigma ^{-}\rightarrow \Sigma ^{0})}
&=&-5r+5s-18x-6y+2z-2p,  \nonumber \\
\left( {g_{1}}/{f_{1}}\right) ^{(\Xi ^{-}\rightarrow \Xi ^{0})}
&=&4r+8s+8x+8y-4z-8q,  \nonumber \\
\left( {g_{1}}/{f_{1}}\right) ^{(\Xi ^{0}\rightarrow \Sigma ^{+})}
&=&-14r+2s+22x+10y+2z+2p-4q.  \label{Eq:semilep}
\end{eqnarray}

The U(3) axial-vector constants $g_{A}^{(0,3,8)}$ can be also expressed in
terms of the new set of parameters Eq.(\ref{Eq:newp}). For the triplet 
ones\footnote{%
Triplet $g^{(3)}$'s are proportional to $I_{3}$, formulae in Eq.(\ref
{Eq:triplet}) correspond to the highest isospin state} we have: 
\begin{eqnarray}
g_{\text{A}}^{(3)}(\text{p}) &=&-14r+2s-44x-20y-4z-4p+8q,  \nonumber \\
g_{\text{A}}^{(3)}(\Lambda ) &=&0,  \nonumber \\
g_{\text{A}}^{(3)}(\Sigma ^{+}) &=&-10r+10s-36x-12y+4z-4p,  \nonumber \\
g_{\text{A}}^{(3)}(\Xi ^{0}) &=&4r+8s+8x+8y-4z-8q,  \label{Eq:triplet}
\end{eqnarray}
and for the octet ones, we get: 
\begin{eqnarray}
g_{\text{A}}^{(8)}(\text{p}) &=&\sqrt{3}(-2r+6s+12x+4p+24q),  \nonumber \\
g_{\text{A}}^{(8)}(\Lambda ) &=&\sqrt{3}(6r+2s-36x+36q),  \nonumber \\
g_{\text{A}}^{(8)}(\Sigma ^{+}) &=&\sqrt{3}(-6r-2s+20x+8y+4p+16q),  \nonumber
\\
g_{\text{A}}^{(8)}(\Xi ^{0}) &=&\sqrt{3}(8r-4s-24x-12y+24q).
\label{Eq:octet}
\end{eqnarray}

As already explained in the Introduction the model provides a link between
the octet currents and the singlet axial current. For the singlet
axial-vector constants, we have: 
\begin{eqnarray}
g_{\text{A}}^{(0)}(\text{p}) &=&60s-18y+6z,  \nonumber \\
g_{\text{A}}^{(0)}(\Lambda ) &=&60s+54y-18z,  \nonumber \\
g_{\text{A}}^{(0)}(\Sigma ) &=&60s-54y+18z,  \nonumber \\
g_{\text{A}}^{(0)}(\Xi ) &=&60s+72y-24z,  \label{Eq:sing}
\end{eqnarray}

Let us note that by redefinition of $q$ and $x$ we can get rid of the
variable $p$: 
\begin{equation}
x^{\prime }=x-\frac{1}{9}p,\;\;\;\;q^{\prime }=q-\frac{1}{9}p.
\end{equation}

In the chiral limit parameters $x$, $y$, $z$, $p$ and $q$ vanish and we
recover the SU(3) symmetric relations from Section II with 
\begin{equation}
D=-3\,s-9\,r,\;F=5\,s-5\,r,
\end{equation}
from which Eq.(\ref{ga0}) follows.

\section{The SU(3) symmetry breaking}

We fix the newly-defined set of parameters from the experimental data of
semileptonic decays. Their numerical values are given in Table III. We do not
quote the experimental errors on these parameters, since they are highly
correlated and cannot be used directly to calculate the errors of the
physical quantities of interest. Instead, we expressed all observables
directly in terms of the $A_{i}$'s. This is, however, not enough since, as
in the chiral limit, the extra input is needed.

At this point a necessity of a complete description of the symmetry breaking
is clearly seen. The strange quark mass causes all SU(3) symmetry relations (%
\ref{Eq:SU3}) to break. So in principle one needs one extra experimental
input for each isospin multiplet. Let us first discuss the case of the
nucleon first.

\subsection{Spin content of the nucleon}

We shall repeat here the analysis of Section II, however, with
the symmetry breaking taken into account. Again four different choices for an 
additional input will be considered: 1) $\Delta \Sigma _{{\rm {p}}}=1$, 2) $%
\Delta s_{{\rm {p}}}=0$, 3) $\Gamma _{{\rm {p}}}=2.56$ and 4) the $\chi $QSM
formulae (\ref{Eq:sing}) for $g_{{\rm {A}}}^{(0)}$. 
The results are summarized in Table II. 
It can be immediately seen that the first two possibilities are in
contradiction with experimental data on $\Gamma _{{\rm {p}}}$ and $\Gamma _{%
{\rm {n}}}$. On the other hand, if we use the experimental value of $\Gamma _{%
{\rm {p}}}$ as an additional input (but no model formula (\ref{Eq:sing}) for 
$g_{{\rm {A}}}^{(0)}$), or alternatively the $\chi$QSM prediction 
for  $g_{{\rm {A}}}^{(0)}$, the results are almost indistinguishable. 
This gives a numerical support for the 
correctness of the $\chi$QSM formula for the axial-vector singlet current 
with the SU(3) symmetry breaking.

Of course the results of Table II have to be taken with a bit of care
because of large experimental errors which are not displayed. As we have
argued in Ref.\cite{KimPraGo2}, one could still accommodate 
$\Delta s_{{\rm {p}}}=0$ due to the large errors of $\Xi $ decays.
We shall come back to this point in the following.

\subsection{Numerical results}

It the present Section we shall present the numerical results of our
analysis based on the Chiral Quark Soliton Model with the SU(3) symmetry 
breaking. Our strategy is very
simple: using model parametrization (\ref{Eq:semilep})
we expressed $\Delta q$'s and $\Delta \Sigma $'s in terms of the six known 
weak semileptonic decays. Errors are added in
quadrature. The numerical results are summarized in Table IV and in Figures
2 -- 9. To guide an eye it is convenient to restore the linear $m_{{\rm {s}}}
$ dependence for the quark densities in the following way: 
\begin{equation}
\Delta q=\Delta q^{(0)}+\frac{m_{{\rm {s}}}}{180\,{\rm {MeV}}}\left( \Delta
q-\Delta q^{(0)}\right) , 
\label{msdep}
\end{equation}
and similarly for $\Delta \Sigma $. This is possible because our chiral
parameters $r$ and $s$ do not need to be refitted as the symmetry breaking
corrections are included. In order to display the errors which come from the
experimental errors of the weak decays, at both ends of each figure we
also plot the theoretical predictions as black dots together with the error 
bars.

Let us first comment on the results on $\Gamma _{{\rm {p}}}$ and $\Gamma _{%
{\rm {n}}}$. We see from Table III that the experimental values are quite
well reproduced by the model, provided the $m_{{\rm {s}}}$ corrections are
included. In the symmetry limit their values are way off from the
experimental data.

Next, let us observe that the singlet axial-vector current couplings 
$g_{\rm {A}}^{(0)}$ split when the symmetry breaking is switched on. This 
is due to the term proportional to $D_{83}^{(8)}$ in Eq.(\ref{Eq:singlet}). 
 This splitting is depicted in Fig.2.  We see that 
$\Delta \Sigma _{{\rm {p}}}$ shows the weakest $m_{{\rm {s}}}$ dependence, 
whereas $\Delta \Sigma _{\Lambda }$ and $\Delta \Sigma _{\Xi }$ depend quite 
strongly on $m_{{\rm {s}}}$. Large error bars for these quantities are due 
almost entirely to the large errors of $\Xi $ decays $A_{5}$ and $A_{6}$. 
It is however evident from Fig. 2 that $\Lambda $ and $\Xi $ are much closer 
to the nonrelativistic limit than p and $\Sigma $.

In Figs.3 -- 6 we plot $\Delta q$ for the nucleon, $\Lambda $, $\Sigma $ and $%
\Xi $ respectively. We see that in all 4 cases $\Delta s$ rises relatively 
strongly 
with $m_{{\rm {s}}}$. It is therefore not justified to extract the
strange quark polarization assuming the exact SU(3) symmetry. Unfortunately, $%
\Delta s$'s have also the largest error coming, as in the case of $\Delta
\Sigma $, almost entirely from the errors of $\Xi $ decays.

In Figs.7--9 we examine the breaking of the SU(3) relations given by Eqs.(\ref
{Eq:SU3}). Interestingly we find that there is an approximate equality
between $\Delta u_{{\rm {p}}}$ and $\Delta u_{\Sigma ^{+}}$ for all values
of $m_{{\rm {s}}}$.

\section{Summary and conclusions}

In the analysis of the polarized structure function $g_1$ of the proton and
neutron one has to take an additional input from the low energy hyperon decays.
Customarily the SU(3) symmetry for these decays is assumed. However, if
one takes all possible combinations of the low energy decays the
resulting $\Delta\Sigma$ can take any value between 0.02 and 0.30. As
depicted in Fig.1 this range is further increased if the errors coming
from the experimental error bars of the semileptonic decays are properly
included. This observation implies that the SU(3) symmetry breaking
plays an essential role in extracting $\Delta\Sigma$ from the
experimental data. It was therefore the aim of this paper to study the
influence of the symmetry breaking on the determination of $\Delta\Sigma$
and $\Delta s$ for the octet baryons in a consistent way.

For this purpose we have performed the "{\em model-independent}" 
analysis based on the algebraic structure of the Chiral Quark Soliton 
Model.  In this approach, one makes merely use of the algebraical structure of 
the model, treating the dynamical quantities which are in principle 
calculable in the model as free parameters. Model predictions of the 
axial-vector properties of the octet
baryons have been already calculated elswhere~\cite{KimPoPraGo}. There
are two model ingredients which are of importance. The first one is  
the model formula for the octet axial-vector currents which have been 
derived in the linear order in $m_{\rm s}$ and $1/N_{\rm c}$. Our formulae 
here have the same algebraical structure as in the large $N_{\rm c}$ QCD 
\cite{Man}, and therefore they are more general than the model itself. 
Secondly, unlike in QCD, the model provides a link between the octet  
axial-vector currents and the singlet axial-vector current.  This 
connection is a truly model-dependent ingredient, however, we have given 
the arguments in favor of Eq.(\ref{Eq:singlet}), based on the fact that 
apart from the general success of the $\chi$QSM in reproducing 
form factors and parton distributions, 
in the limit of the small soliton it properly reduces to the Nonrelativistic 
Quark Model prediction, and in the limit of the large soliton 
it reproduces the Skyrme Model prediction for
$\Delta\Sigma$.  Similarly, in Ref.\cite{WeiSig} the argument has been
given that Eq.(\ref{ga0}) naturally emerges in the limit of the large 
$m_{\rm s}$, where the SU(3) flavor symmetry reduces to the SU(2) one.  
The numerical analysis of Section V.A provides a further support for the 
model formula for $\Delta\Sigma$.

We have presented two parametrizations of all available semileptonic decays. 
The first one is obtained assuming the SU(3) symmetry, however the two reduced
matrix elements $F$ and $D$ were extracted from the combinations of the
semileptonic decays which are free of the $m_{\rm s}$ corrections 
(\ref{DFexp}), rather than from the neutron and $\Sigma^-$ decays alone.
The second one is obtained by fitting all 6 measured semileptonic decays
in terms of 6 free parameters defined in Eqs.(\ref{Eq:newp},\ref{Eq:semilep}).
The difference between the two fits, as seen from Table III, is rather small, 
except perhaps for the $\Sigma^- \rightarrow {\rm n}$ decay. Despite the
fact that the symmetry breaking for the semileptonic decays themselves is not
strong, other quantities like $\Delta s$ and $\Delta \Sigma$ are much
more affected by taking into account the effects of the non-zero strange
quark mass. This is clearly shown in Figs.2--9. 

Whether this sensitivity is a sign of the breakdown of the perturbative 
approach to the strangenes, as it was recently suggested in 
Ref.\cite{WeiSig}, is hard to say, since our anaysis suffers from large 
errors which are mainly due to the experimental errors of the $\Xi^-$ decays. 
It is therefore of utmost importance to measure these two decays with 
the precision comparable to the other four decays.  One should bare in mind 
that this is one of a few cases, where the low energy data have an 
important impact on our understanding of the high energy scattering.
Given the theoretical implications of these experiments as far as the
role of the axial anomaly and the gloun polarization is concerned 
\cite{Cheng,Goto,AEL},
one should make it clear how important the new measurements of the 
$\Xi^-$ decays would be. This is perhaps the most important message of
our analysis.

\section*{Acknowledgments}
The work of H.-Ch.K has been supported by the Korean Physical Society.
The work of M.P. has been supported by Polish KBN Grant PB~2~P03B~019~17.
The work of K.G. has been supported by the BMBF, the DFG, and the
COSY--Project(J\"{u}lich). 

\newpage


\newpage

\begin{table}
\caption{
The results for $\Delta q$'s, $\Delta\Sigma$ and $\Gamma_{\rm p}$ for
various phenomenological inputs (denoted by a $\ast$) in the case of 
the exact SU(3) symmetry.
}
\begin{tabular}{crrrrrrr}
~ & NRQM & \multicolumn{2}{c}{Ellis \& Jaffe} & \multicolumn{2}{c}{$\Gamma
_{{\rm {p}}}=2.56$} & \multicolumn{2}{c}{$\chi $QSM} \\ \hline
~ &  & $A_{1},\,A_{4}$ & average & $A_{1},\,A_{4}$ & average & $A_{1},\,A_{4}
$ ~ & average \\ \hline\hline
$D^{\ast }$ & $1$ & $0.80$ & $0.77$ & $0.80$ & $0.77$ & $0.80$ & $0.77$ \\ 
\hline
$F^{\ast }$ & $2/3$ & $0.46$ & $0.50$ & $0.46$ & $0.50$ & $0.46$ & $0.50$ \\ 
\hline\hline
$\Delta u_{{\rm {p}}}$ & $4/3$ & $0.92$ & $1.00$ & $0.79$ & $0.81$ & $0.77$
& $0.98$ \\ \hline
$\Delta d_{{\rm {p}}}$ & $-1/3$ & $-0.34$ & $-0.27$ & $-0.47$ & $-0.47$ & $%
-0.49$ & $-0.29$ \\ \hline
$\Delta s_{{\rm {p}}}$ & $0$ & $0^{\ast }$ & $0^{\ast }$ & $-0.13$ & $-0.20$
& $-0.15$ & $-0.02$ \\ \hline
$\Delta u_{\Lambda }$ & $0$ & $-0.07$ & $-0.01$ & $-0.20$ & $-0.21$ & $-0.22$
& $-0.03$ \\ \hline
$\Delta s_{\Lambda }$ & $1$ & $0.76$ & $0.76$ & $0.60$ & $0.56$ & $0.58$ & $%
0.74$ \\ \hline
$\Delta u_{\Sigma ^{0}}$ & $2/3$ & $0.50$ & $0.50$ & $0.33$ & $0.30$ & $0.31$
& $0.48$ \\ \hline
$\Delta \Sigma $ & $1^{\ast }$ & $0.58$ & $0.74$ & $0.19$ & $0.14$ & $0.13$
& $0.68$ \\ \hline
$\Gamma _{{\rm {p}}}$ & $5$ & $3.34$ & $3.75$ & $2.56^{\ast }$ & $2.56^{\ast
} $ & $2.44$ & $3.63$ 
\end{tabular}
\end{table}

\vspace{0.5cm}

\begin{table}
\caption{
$\Delta s_{\rm p}$, $\Delta\Sigma_{\rm p}$ and $\Gamma_{\rm p,n}$ for
various phenomenological inputs (denoted by a $\ast$) in the case of 
the broken SU(3) symmetry.
}
\begin{tabular}{crrrr}
& NRQM & Ellis \& Jaffe & $\Gamma _{{\rm {p}}}=2.56$ & $\chi $QSM \\ 
\hline
$\Delta \Sigma _{{\rm {p}}}$ & $1^{\ast }$ & $-0.47$ & $0.56$ & $0.51$ \\ 
\hline
$\Delta s_{{\rm {p}}}$ & $0.49$ & $0^{\ast }$ & $0.31$ & $0.32$ \\ \hline
$\Gamma _{{\rm {p}}}$ & $3.65$ & $0.71$ & $2.56^{\ast }$ & $2.67$ \\ \hline
$\Gamma _{{\rm {n}}}$ & $-0.12$ & $-3.06$ & $-1.21$ & $-1.10$ 
\end{tabular}
\end{table}

\vspace{0.5cm}

\begin{table}
\caption{Model parameters $r \ldots q'$ extracted from the data
together with the predictions for the semileptonic decays and
$\Gamma_{\rm p,n}$ in the case of the exact SU(3) and broken SU(3). 
Results for $A_i$'s with $m_{{\rm s}}$ corrections correspond to the
experimental data \protect\cite{PDG96}.
}
\begin{tabular}{ccrr}
~&  & exact SU(3) & broken SU(3) \\ \hline
& $r$ & $-0.0892 $ & $-0.0892 $ \\ 
& $s$ & $0.0113 $ & $0.0113 $ \\ 
& $x^{\prime}$ & $0~~~~ $ & $-0.0055 $ \\ 
& $y$ & $0~~~~ $ & $0.0080 $ \\ 
& $z$ & $0~~~~ $ & $-0.0038 $ \\ 
& $q^{\prime}$ & $0~~~~ $ & $-0.0140 $ \\ \hline
$A_1$ & $\left({g_1}/{f_1}\right)^{n\rightarrow p}$ & $1.271\pm 0.11$ & $%
1.2573\pm 0.0028$ \\ 
$A_2$ & $\left({g_1}/{f_1}\right)^{\Sigma^+\rightarrow \Lambda}$ & $0.769\pm
0.04$ & $0.742 \pm 0.018 $ \\ 
$A_3$ & $\left({g_1}/{f_1}\right)^{\Lambda\rightarrow p}$ & $0.758\pm 0.08$
& $0.718 \pm 0.015 $ \\ 
$A_4$ & $\left({g_1}/{f_1}\right)^{\Sigma^-\rightarrow n}$ & $-0.267\pm0.04$
& $-0.340 \pm 0.017 $ \\ 
$A_5$ & $\left({g_1}/{f_1}\right)^{\Xi^-\rightarrow \Lambda}$ & $0.246\pm
0.07$ & $0.25 \pm 0.05 $ \\ 
$A_6$ & $\left({g_1}/{f_1}\right)^{\Xi^-\rightarrow \Sigma^0}$ & $1.271\pm
0.11$ & $1.278 \pm 0.158 $ \\ \hline
~ & $\Gamma_p$ & $3.63 \pm 1.12$ & $2.67 \pm 0.33 $ \\ 
~ & $\Gamma_n$ & $-0.19 \pm 0.84$ & $-1.10 \pm 0.33 $ 
\end{tabular}
\end{table}

\vspace{0.5cm}

\begin{table}
\caption{Integrated polarized quark densities for various baryons. }
\begin{tabular}{lrr}
& exact SU(3) & broken SU(3)\\ \hline
$\Delta u_{{\rm p}}=\Delta d_{{\rm n}}$ & $0.98\pm 0.23$ & $0.72 \pm 0.07$
\\ 
$\Delta d_{{\rm p}}=\Delta u_{{\rm n}}$ & $-0.29\pm 0.13$ & $-0.54\pm 0.07$
\\ 
$\Delta s_{{\rm p}}=\Delta s_{{\rm n}}$ & $-0.02\pm 0.09$ & $0.33\pm 0.51$
\\ \hline
$\Delta u_\Lambda=\Delta d_\Lambda$ & $-0.03 \pm 0.14$ & $-0.02 \pm 0.17$ \\ 
$\Delta s_\Lambda$ & $0.74 \pm 0.17$ & $1.21 \pm 0.54$ \\ \hline
$\Delta u_{\Sigma^+}=\Delta d_{\Sigma^-}$ & $0.98\pm 0.23$ & $0.73 \pm 0.17$
\\ 
$\Delta d_{\Sigma^+}=\Delta u_{\Sigma^-}$ & $-0.02\pm 0.09$ & $-0.37\pm 0.19$
\\ 
$\Delta s_{\Sigma^+}=\Delta s_{\Sigma^-}=\Delta s_{\Sigma^0}$ & $-0.29\pm
0.13$ & $-0.18\pm 0.39$ \\ 
$\Delta u_{\Sigma^0}=\Delta d_{\Sigma^0}$ & $0.48\pm 0.16$ & $0.18\pm 0.08$
\\ \hline
$\Delta u_{\Xi^0}=\Delta d_{\Xi^-}$ & $-0.29 \pm 0.13$ & $-0.14 \pm 0.21$ \\ 
$\Delta d_{\Xi^0}=\Delta u_{\Xi^-}$ & $-0.02 \pm 0.09$ & $0.02 \pm 0.16$ \\ 
$\Delta s_{\Xi^0}=\Delta s_{\Xi^-}$ & $0.98 \pm 0.23$ & $1.50 \pm 0.60$%
\end{tabular}
\end{table}

\newpage

\begin{figure}
\centerline{\epsfysize=3.5in\epsffile{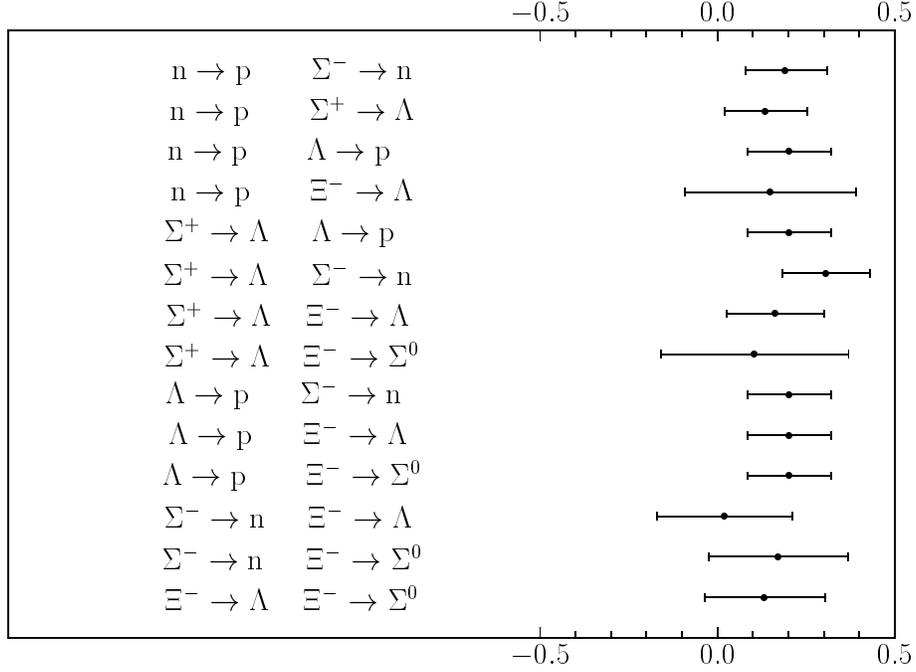}} \vskip4pt
\caption{
$\Delta \Sigma _{{\rm {p}}}$ with $\Gamma_{\rm p}$ and different 
semileptonic decays taken as an input in the SU(3) symmetry limit.}
\end{figure}

\begin{figure}
\centerline{\epsfysize=2.7in\epsffile{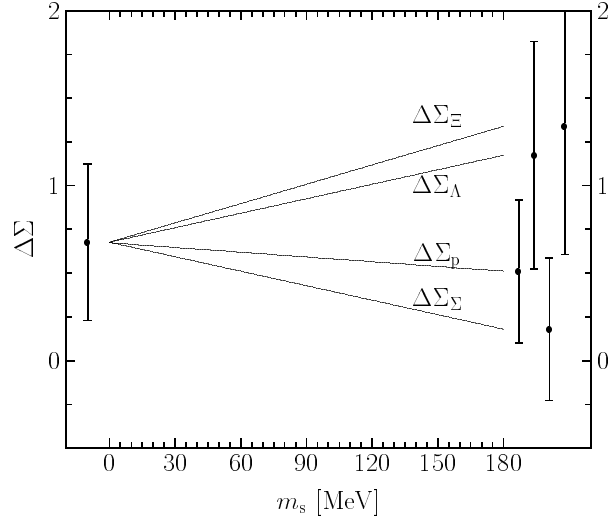}} \vskip4pt
\caption{
$\Delta \Sigma _{{\rm {B}}}$ with and without SU(3) symmetry 
breaking.
In the $\chi$QSM with $m_{\rm s}$ dependence restored according to 
Eq.(\ref{msdep}). Black dots denote model predictions (same as 
lines) 
with errors coming from the experimental errors of the semileptonic 
weak decays. }
\end{figure}

\vspace{0.5cm}

\begin{figure}
\centerline{\epsfysize=2.7in \epsffile{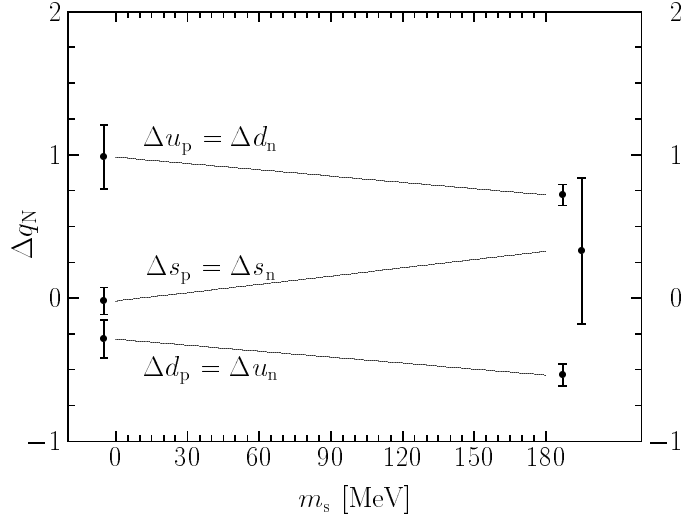}} \vskip4pt 
\caption{
$\Delta q$' for the nucleon; dots and error bars have the same 
meaning as in Fig.2.}
\end{figure}

\vspace{0.5cm}

\begin{figure}
\centerline{\epsfysize=2.7in \epsffile{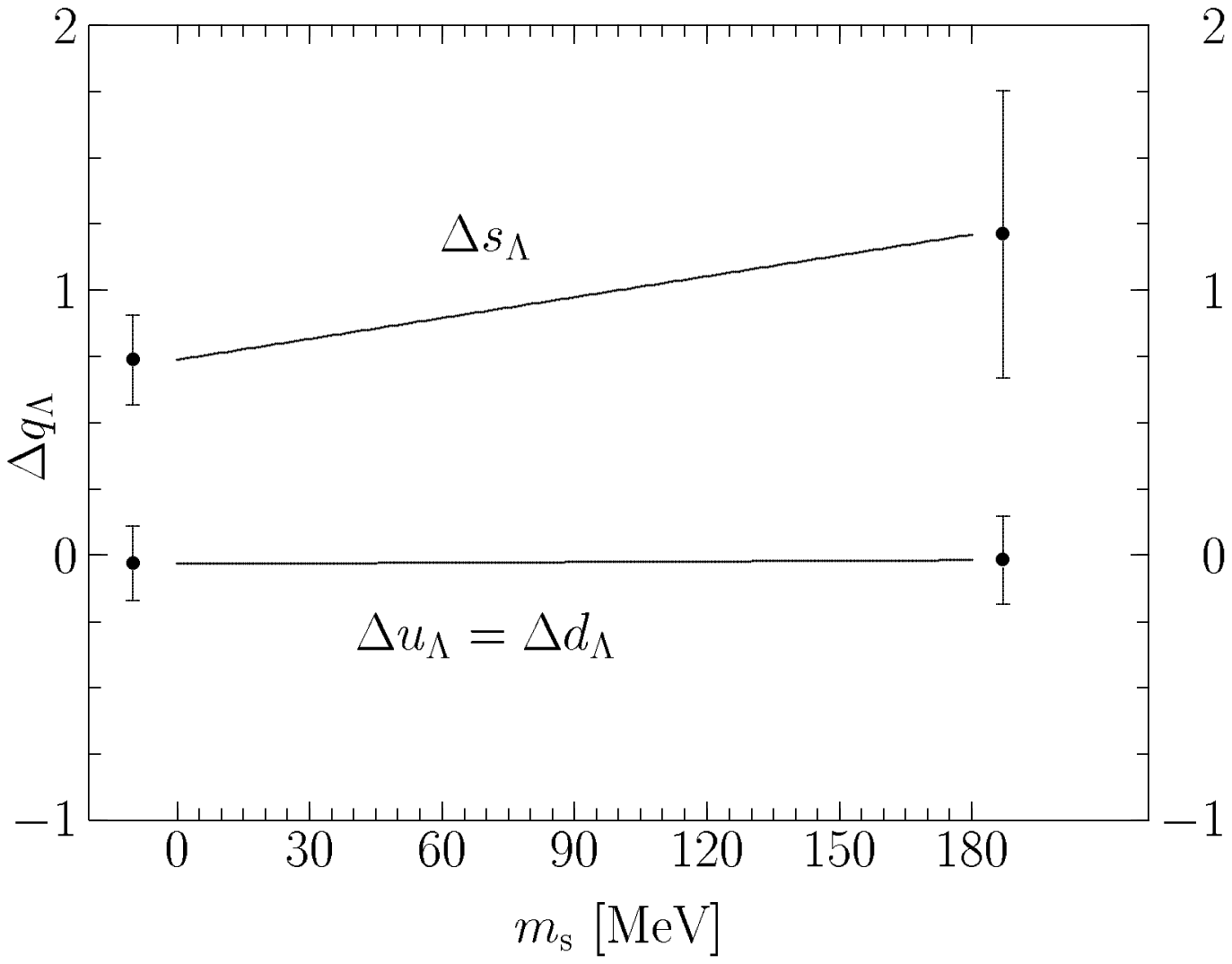}} \vskip4pt 
\caption{
$\Delta q$' for the $\Lambda$ ; dots and error bars have the same 
meaning as in Fig.2. }
\end{figure}

\vspace{0.5cm}

\begin{figure}
\centerline{\epsfysize=2.7in\epsffile{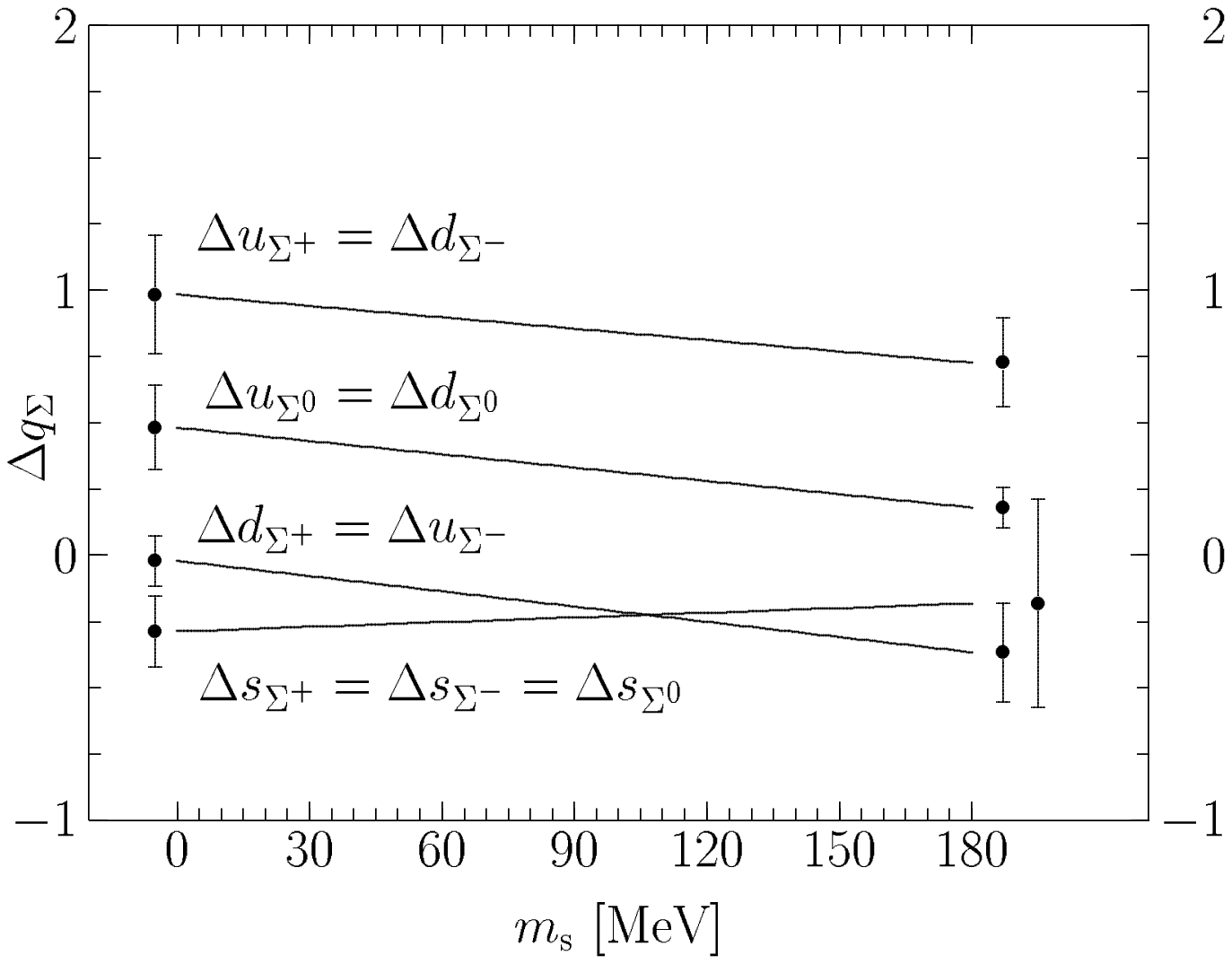}} \vskip4pt 
\caption{
$\Delta q$' for the $\Sigma$ ; dots and error bars have the same 
meaning as in Fig.2. }
\end{figure}

\vspace{0.5cm}

\begin{figure}
\centerline{\epsfysize=2.7in\epsffile{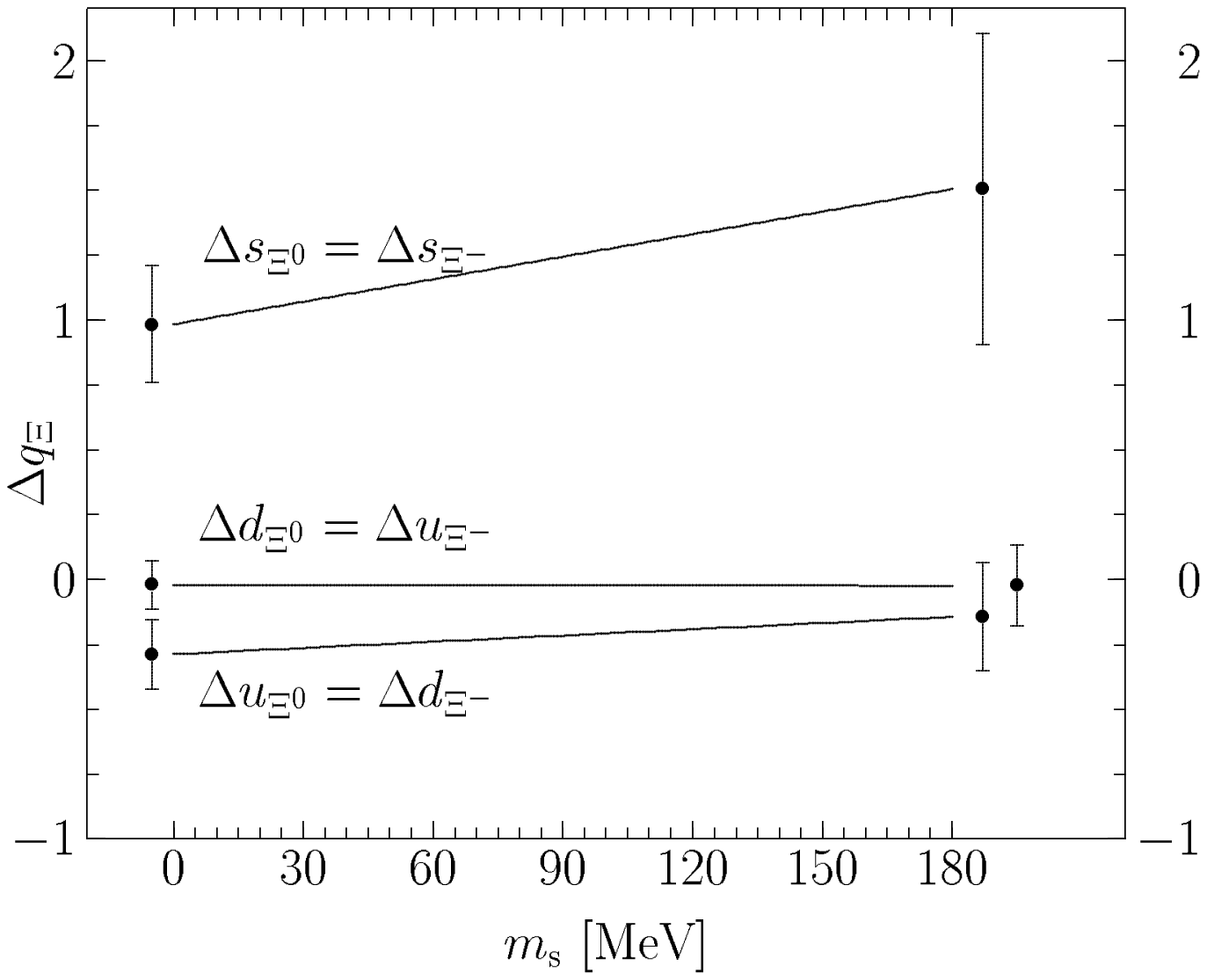}} \vskip4pt 
\caption{
$\Delta q$' for the $\Xi$ ; dots and error bars have the same 
meaning as in Fig.2.}
\end{figure}

\vspace{0.5cm}

\begin{figure}
\centerline{\epsfysize=2.7in\epsffile{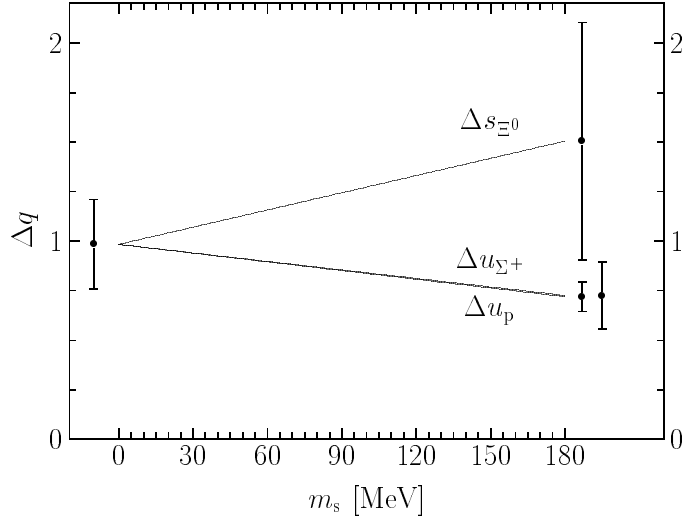}} \vskip4pt 
\caption{
Breaking of the first SU(3) relation of Eq.(\ref{Eq:SU3}); 
dots and error bars have the same 
meaning as in Fig.2. The point with large error bar at the lower curves 
corresponds to $\Sigma^+$.}
\end{figure}

\vspace{0.5cm}

\begin{figure}
\centerline{\epsfysize=2.7in\epsffile{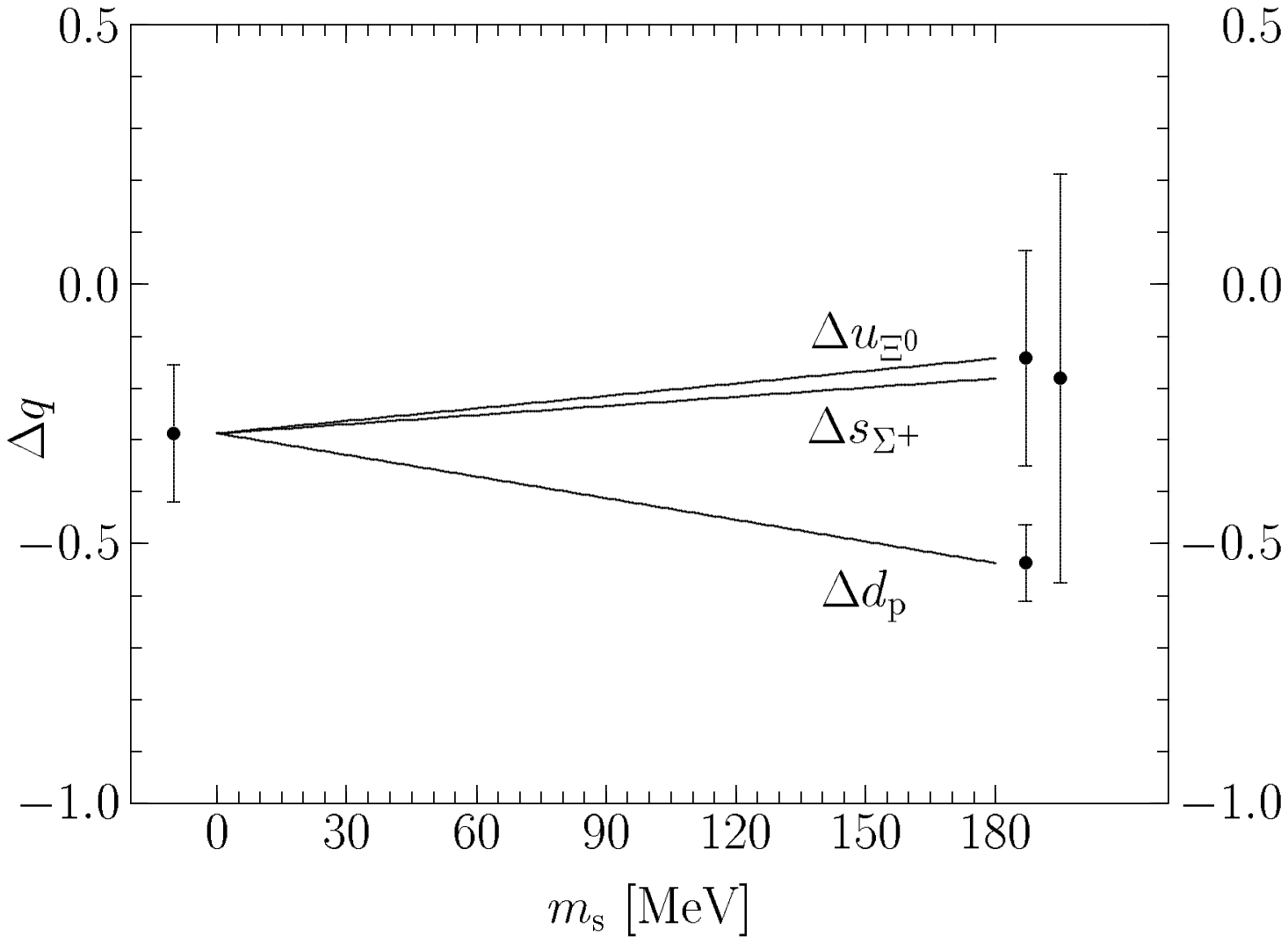}} \vskip4pt 
\caption{
Breaking of the second SU(3) relation of Eq.(\ref{Eq:SU3}); 
dots and error bars have the same 
meaning as in Fig.2. }
\end{figure}

\vspace{0.5cm}

\begin{figure}
\centerline{\epsfysize=2.7in\epsffile{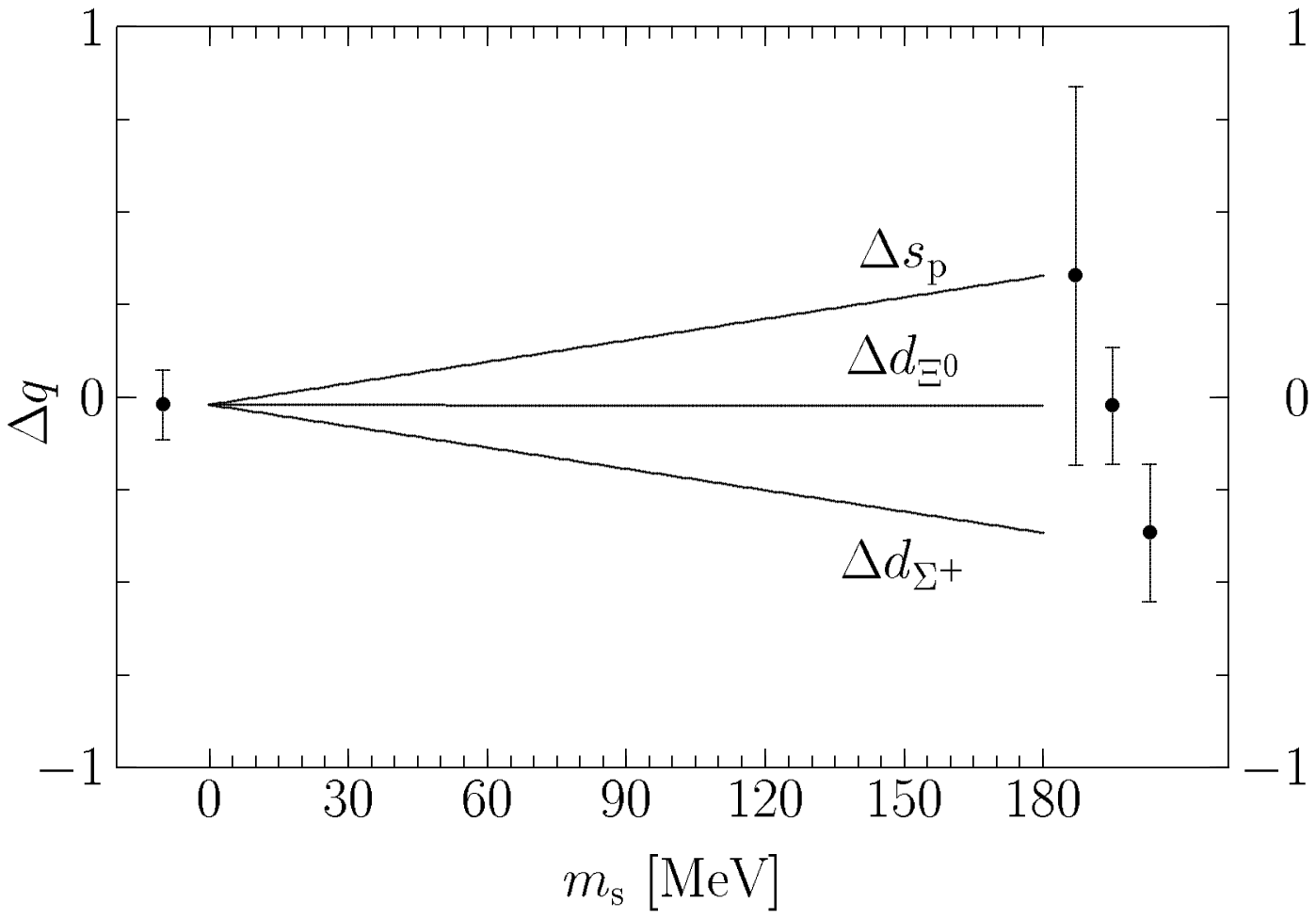}} \vskip4pt 
\caption{
Breaking of the third SU(3) relation of Eq.(\ref{Eq:SU3}); 
dots and error bars have the same 
meaning as in Fig.2. }
\end{figure}

\end{document}